# Surface energy engineering of graphene

Young Jun Shin,[†,‡] Yingying Wang,[§] Han Huang,[⊥] Gopinadhan Kalon,[†,‡] Andrew Thye Shen Wee,[⊥] Zexiang Shen,[§] Charanjit Singh Bhatia,[†,∥] and Hyunsoo Yang[*,†,‡]

[†]Department of Electrical Engineering and Computer Engineering, National University of Singapore Singapore 117576, [‡]NanoCore, National University of Singapore, Singapore 117576, [§]Division of Physics and Applied Physics, School of Physical and Mathematical Sciences, Nanyang Technological University, Singapore 637371, [⊥]Department of Physics, National University of Singapore, Singapore 117542, and [∥]Institute of Materials Research and Engineering, Singapore 117602

[*]Corresponding author's email: eleyang@nus.edu.sg

**Abstract:** Contact angle goniometry is conducted for epitaxial graphene on SiC. Although only a single layer of epitaxial graphene exists on SiC, the contact angle drastically changes from 69° on SiC substrates to 92° with graphene. It is found that there is no thickness dependence of the contact angle from the measurements of single, bi, and multi layer graphene and highly ordered pyrolytic graphite (HOPG). After graphene is treated with oxygen plasma, the level of damage is investigated by Raman spectroscopy and correlation between the level of disorder and wettability is reported. By using low power oxygen plasma treatment, the wettability of graphene is improved without additional damage, which can solve the adhesion issues involved in the fabrication of graphene devices.

## Introduction

Graphene has attracted much attention due to its superior characteristics such as very high mobility and massless Dirac fermions character.[1-6] Due to its exceptional properties, wide applications have been investigated including field effect transistors (FET) and logic devices.[7-11] Recently a few methods for synthesizing large area graphene, an important step towards electronic applications, were proposed[6, 12-14] making graphene a promising material for post silicon era. In order to fabricate useful devices, understanding the surface property of graphene is very important since the contact deposition is critical for the device performance and functionality and the contact deposition often fails due to graphene's hydrophobic nature. However, there have been few studies investigating the surface property of graphene as opposed to many reports on the electrical properties, due to limited size of the single layer of epitaxial graphene (EG). Here we investigate the wettability of graphene on SiC by contact angle measurements. The dependence of the wettability on the number of graphene layers was also investigated. By treating graphene with oxygen plasma, disorder or defect was introduced on graphene and the level of disorder was determined by Raman spectroscopy.[15,16] The correlation between the level of disorder and contact angle of graphene provides more insight on the physical meaning of D band in Raman spectroscopy.[17] We also propose a method to improve the adhesion between metal contacts and the graphene surface by controlling the surface property which will introduce little or no damage to graphene.

## Experimental Section

EG on SiC was prepared by annealing chemically etched (10% HF solution) n-type Si-terminated 6$H$-SiC (0001) samples at 850 °C under a silicon flux for 2 minutes in ultra-high vacuum (UHV), resulting in a Si-rich 3×3-reconstructed surface, and subsequently annealing at a higher temperature (> 1200 °C) in the absence of the silicon flux. The thickness of EG films can be controlled by the annealing temperature and time followed by slow cooling to the room temperature, allowing the preparation of samples with EG thicknesses ranging from one to three and more layers.[12,18] The temperature of the samples was measured by an optical pyrometer. The mechanically cleaved graphene (MCG) was prepared by micromechanical exfoliation and transferred to Si substrates which was covered by a layer of 300nm $SiO_2$.[19] The quality of graphene was examined by scanning tunneling microscopy (STM), atomic force microscopy (AFM), and Raman spectroscopy. AFM imaging of graphene has been carried out in the contact mode and with environmental chamber which can remove moisture by controlling the environment with silica gel. The Raman spectra were obtained using 532 nm (2.33 eV) laser lines as the excitation source and laser power on the sample was below 0.5 mW to avoid laser induced heating. An objective lens with a magnification of 100X and a numerical aperture of 0.95 has been used and the focused laser spot size is ~ 500nm in diameter.[15] The measurement of the contact angle of graphene has been carried out in ambient conditions. 0.5 μL of de-ionized water droplet has been released onto graphene surface from a syringe needle. The image of the liquid droplet was obtained in real time by using a CCD camera. A tangent line has been drawn onto the droplet from the droplet-graphene interface in the image, and the angle between the tangent line and the base line indicates the contact angle of the solid and liquid interface. The accuracy of the contact angle measurements is less than 1°. The contact angle data of twenty measurements per sample were averaged.

**Results and discussion**

Figure 1a,b shows *in-situ* STM images of single and bi-layer EG on *6H*-SiC. Honeycomb structure is clearly observed in the STM images. Detail information on the preparation and confirmation of graphene on SiC is described elsewhere.[12, 18] The AFM image in Figure 1c clearly shows that the atomic arrangements are repeated in a regular fashion with possible distortions due to the mechanical and thermal drift. Figure 1d shows the Raman spectra of graphene on SiC and pristine SiC substrate. The appearance of the in-plane vibrational G band (1597 $cm^{-1}$) and two phonon 2D band (2715 $cm^{-1}$) after decomposing Si from SiC indicates that graphene has been formed on SiC.[15]

Figure 2 shows the images of water droplets on SiC substrate, highly ordered pyrolytic graphite (HOPG), single layer EG, and plasma etched graphene on SiC. Although only one monolayer of graphene exists on top of SiC substrate, there is a drastic change in the contact angle of the water droplet with graphene (92.5°) on SiC compared to that of pristine SiC (69.3°). The contact angle of freshly cleaved HOPG (91°) shows that it is hydrophobic similar to graphene on SiC. After completely etching monolayer of graphene by oxygen plasma at 10 W for 2 min, the contact angle (70°) is similar to that of a SiC substrate.

Properties of graphene are sensitive to the number of layers. Different characteristics such as the electrical and mechanical properties of mono-layer, bi-layer, and tri-layer have been investigated by different groups.[20-22] In order to determine the dependence of wettability of graphene on the layer thickness, we measured the contact angle as a function of number of layers and the result is summarized in Table 1. By investigating the measurement result we concluded that the wettability of graphene is independent of the thickness. Wang *et al.* reported a contact angle of 120° on graphene films[23] which is quite different from our result. This difference can be attributed to their rough surface, a result of integrating many micro scale and different layer of graphene flakes. This is consistent with the fact that a hydrophobic surface becomes more hydrophobic when microstructured.[24] The change of the contact angle due to surface roughness can be described by $\cos(\theta_W) = R_w \cos(\theta_0)$, where $\theta_W$ is the apparent contact angle, $R_w$ is the surface roughness factor, and $\theta_0$ is the contact angle in the Young's mode.[25]

In order to investigate the difference of wettability between epitaxial graphene and disordered graphene, we introduced damage and defects intentionally by oxygen plasma treatment. Figure 3 summarizes the contact angle measurements on graphene for different conditions. After oxygen plasma treatment at 5 W for 15 sec, the surface becomes more hydrophilic with the contact angle changing from 92.5° of pristine graphene to 55.1° in Figure 3a,b. Oxygen plasma could create vacancies, C-H, $sp^3$ sites, or $OH^-$ bonding.[26] However, when the plasma power is relatively small (< 2 W) and the exposure time is less than 1 minute, no detectable defect is present in graphene from the Raman and AFM measurements. Moreover, the contact angle is completely recovered from 42.4° with plasma treatment of 2 W for 45 sec whereby no defect is present to 91.6° by dehydroxylation process after annealing at 300 °C in UHV for 30 minutes.[27] We monitored that the contact angle of graphene was changed to 72.4° after one day of oxygen plasma treatment in Figure 3c. It is well known that most of the hydrophilic characteristic could be originated from hydroxide introduced by oxygen plasma process.[28] After annealing the plasma treated samples at 300 °C in UHV for 30 minutes, contact angle of graphene has been recovered from 72.4° to 87.3° as shown in Figure 3c,d.

It was shown that the contact angle of 92.5° for graphene layer on SiC changes to 70° in Figure 2 when the graphene layer was removed by oxygen plasma. With an oxygen plasma condition of 10 W for 2 minutes it is confirmed that graphene on SiC has been completely removed since the G peak (1597 cm$^{-1}$) and the 2D peak (2715cm$^{-1}$) are not present in the Raman spectra as shown in Figure 4a. It was found that there is a significant difference between the etching rate of MCG and EG.[29] When RF power was 5 W and exposure time was 5 seconds, a more pronounced D band from MCG indicates that MCG is more reactive with oxygen plasma in comparison to EG as shown in Figure 4a,b. Since relatively stronger covalent bond exists between EG and SiC substrate compared to MCG and SiO$_2$ layer, SiC or buffer layer may hold EG more tightly during oxygen plasma treatment.[30,31] This is in line with the previous study in which the authors reported very different etching rate for single and multi layer graphene.[29] Since single layer graphene is bonded to SiO$_2$ more loosely by van der Waals forces as compared to the bonding between graphene layers in mechanically cleaved multi layer graphene, the etching rate of a single layer graphene is faster.[29] In the present study, it took about 25 seconds to completely etch MCG with a RF power of 5W whereas it took 3 minutes to remove EG at the same RF power.

By controlling the exposure time of oxygen plasma, different levels of damage can be introduced on graphene samples. By calculating the integral intensity ratio of the D band to G band, I(D)/I(G), from the Raman spectra, the level of defect on graphene is extracted.[26] It is clear from Figure 5 that the relative intensity of D band to G band increases with increase in exposure time. Figure 5b shows that the contact angle generally decreases with increase in I(D)/I(G) ratio. The result indicates that the defects, which could be a surface dislocation, corrugation, and interaction of graphene with the substrate or vacancies, have increased the polarity of the surface, therefore the surface energy has increased.[15] Further studies are required to better understand the interaction of oxygen plasma with graphene since the role of various defects in controlling the contact angle is not clear.

Understanding the surface characteristics and controlling the wettability of graphene are very important for many applications. In contact deposition on top of graphene, without sufficient understanding of the surface characteristics of graphene, the process is not always guaranteed to be successful. For example, Figure 6a shows the unsuccessful attempt of contact deposition. Most of 5 nm Cr/100 nm Au contact electrodes have been deposited successfully on top of $SiO_2$, whereas part of the electrodes which is located on the graphene side have been peeled off during the lift-off process. As shown in the contact angle experiments, graphene surface is not adhesive (hydrophobic). By oxygen plasma treatment, we can improve the adhesion property of graphene. It should be noted that oxygen plasma exposure time and power should be carefully selected in order to minimize the damage on graphene and achieve good adhesion of the contacts at the same time. In the case of 30 sec etching time, a big difference of the I(D)/I(G) ratio, proportional to the level of damage, between 2W and 5W of RF power is shown in Figure 6b. As we already mentioned, MCG is etched and damaged by oxygen plasma at a faster rate compared to EG at the same power of 2 W. Graphene transforms from hydrophobic to hydrophilic after the oxygen plasma treatment as inferred from the contact angle change from 92° to 10°. For EG, no significant rise of D band has been observed during the entire plasma process indicating no significant damage on graphene when 2 W RF power is used. In contrast, for MCG when exposure time has reached 30 seconds, the D band started rising significantly. This method to control wettability can be combined with the annealing process, which can cure any damage induced by oxygen plasma, thus providing good contact adhesion without compromising the physical properties. Liang *et al*. fabricated graphene transistor without and with oxygen plasma treatment.[32] They reported that oxygen plasma treatment possibly increases the bonding strength even though the dangling bond generated by plasma treatment could degrade mobility. It is consistent with the results of the present study and can be explained to be a result of the improved adhesion due to the presence of hydroxyl group, which increases the polarity of surface. By choosing appropriate power, less than 2 W in the present study, and time of oxygen plasma, high performance graphene devices can be fabricated with good adhesion between graphene and metal contacts as well as minimal damage to graphene.

**Summary**

The wettability of epitaxial graphene on SiC has been studied by contact angle measurements. Mono layer epitaxial graphene showed a hydrophobic characteristic similar to HOPG and no correlation was found between different layers of graphene and wettability. Upon oxygen plasma treatment, defects are introduced into graphene and the level of damage was investigated by Raman spectroscopy. There exists a correlation between the level of defects and the contact angle. As more defects are induced, surface energy of graphene is increased leading to hydrophilic nature. Oxygen plasma treatment with an optimized power and duration has been proposed to control the adhesion property for contact fabrication.

**Figure captions**

**Figure 1.** (a) 2nm × 2nm STM image of single layer graphene on 6*H*-SiC (0001). (b) 8nm × 8nm STM image of bi layer graphene on 6*H*-SiC (0001). (c) AFM image of single layer graphene on 6*H*-SiC (0001). (d) Raman spectra of single layer graphene and SiC substrate.

**Figure 2.** Water droplet on SiC (a), HOPG (b), single layer graphene on SiC (c), and oxygen plasma etched graphene on SiC at 10 W for 2 min (d).

**Figure 3.** Water droplet on graphene before plasma treatment (a), after plasma treatment (5 W, 15 sec) (b), 1 day after $O_2$ plasma treatment (c), and annealed at 300°C in UHV for 30 min (d).

**Figure 4.** (a) Raman spectra of EG without and with plasma treatment. (b) Raman spectra of MCG without and with plasma treatment.

**Figure 5.** (a) Raman spectra of EG treated with 5 W plasma as a function of exposure time. (b) Contact angle versus I(D)/I(G) ratio and I(D)/I(G) ratio versus plasma exposure time.

**Figure 6.** (a) Image of graphene devices when part of the electrodes are peeled off after lift-off process (scale bar: 10 μm, electrodes were supposed to be deposited in the area guided by black line). (b) The $O_2$ plasma exposure time dependence of contact angle and I(D)/I(G) ratio. The plasma power is indicated in brackets.

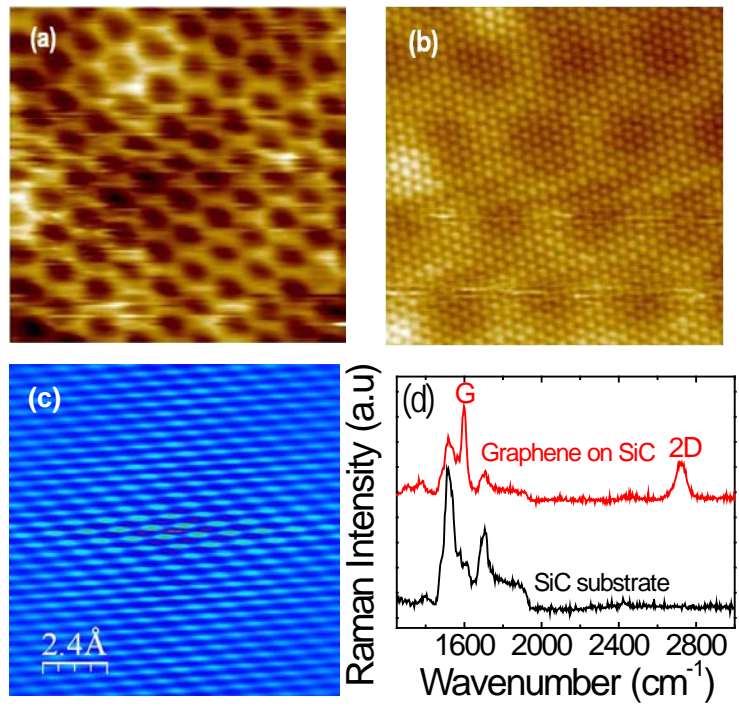

**Figure 1**

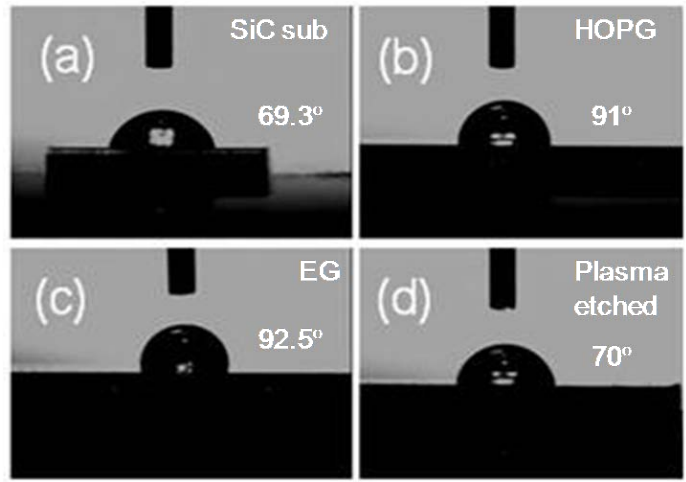

**Figure 2**

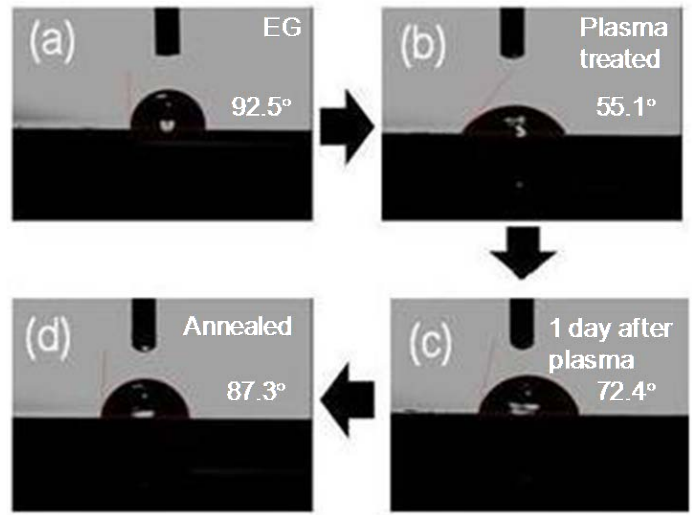

**Figure 3**

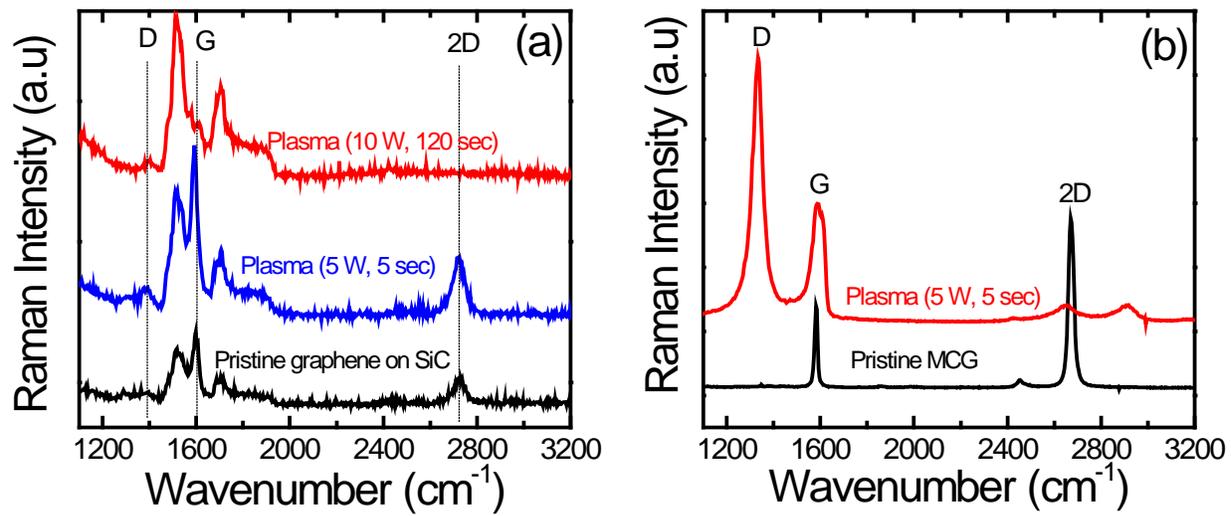

**Figure 4**

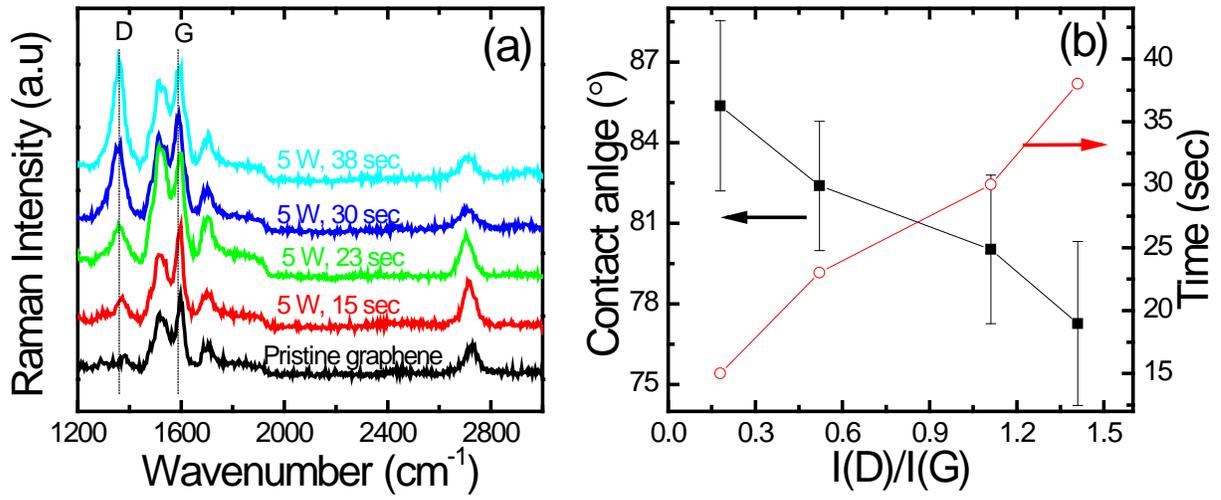

**Figure 5**

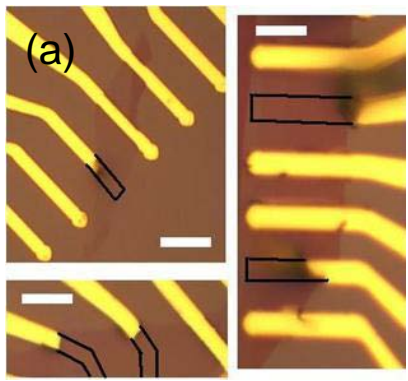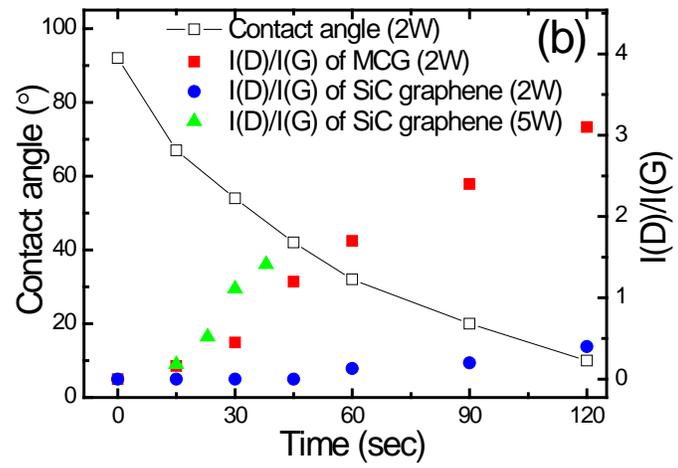

Figure 6

**Table 1.** Averaged contact angle of graphene with different number of layers.

| Number of layer | Contact angle (standard deviation) |
| --- | --- |
| Single | 92.5° (2.9) |
| Bi | 91.9° (3.4) |
| Multi | 92.7° (2.3) |
| HOPG | 91° (1.0) |

**Table of Contents Graphic**

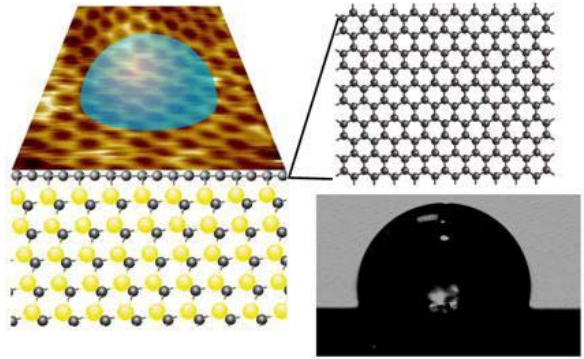